\documentclass[showpacs,amssymb,amsmath,aps,prd,secnumarabic,twocolumn,groupedaddress]{revtex4-1}

\usepackage{graphicx}

\newcommand{\pt}{$p_{\rm T}$}

\begin{document}

\title{Rotation-invariant observables \\ in parity-violating decays of vector
particles to fermion pairs}

\author{Pietro Faccioli$^a$, Carlos Louren\c{c}o$^b$, Jo\~ao
  Seixas$^{a,c}$, and Hermine K. W\"{o}hri$^a$}

\affiliation{$^a$LIP, 1000-149 Lisbon, Portugal\\
$^b$CERN, 1211 Geneva, Switzerland\\
$^c$Physics Department, IST, 1049-001 Lisbon, Portugal}

\date{\today}

\begin{abstract}

The di-fermion angular distribution observed in decays of inclusively produced
vector particles is characterized by two frame-independent observables,
reflecting the average spin-alignment of the
produced particle and the magnitude of parity violation in the decay.
The existence of these observables derives from the rotational properties of
angular momentum eigenstates and is a completely general result, valid for any
$J=1$ state and independent of the production process.
Rotation-invariant formulations of polarization and of the decay
parity-asymmetry can provide more significant measurements than the commonly
used frame-dependent definitions, also improving the quality of the
comparisons between the measurements and the theoretical calculations.

\end{abstract}



\maketitle


\section{Introduction}

Experimental studies of the decay angular distributions of vector
particles represent a detailed way of testing fundamental
theories.
Quarkonium polarization measurements, in particular, are expected to
provide key information for the understanding of quantum
chromodynamics (QCD)~\cite{bib:YellowRep-QWG}, with competing
production mechanisms leading to very different polarization
predictions.
Similarly, studies of the parity violating decays of heavy gauge bosons in a
fermion-antifermion pair, such as $Z / \gamma^{*} \to \ell\ell$ and
$W\to\ell\nu$, provide detailed tests of the electroweak theory,
accurate determinations of the effective weak mixing angle, or lead to the
discovery of new vector bosons coupling to fermion pairs~\cite{bib:newZs}.
The outcome of such studies can be limited --- especially when the production
mechanism of the particles under study is unknown --- by the choice of specific
polarization frames, which may result in misleading interpretations of
polarization measurements, prevent model-independent physical conclusions, or
lead to an effective reduction of the maximally observable parity-violation
asymmetry.

In this paper we show how the rotation-covariance properties of
angular momentum eigenstates imply the existence of a frame-invariant
relation among the parameters characterizing the polar and azimuthal
anisotropies of the di-fermion decay distribution of inclusively
observed vector particles.
This relation has been previously established in the case of parity-conserving
decays~\cite{bib:LTGen} and its advantages in the studies of J/$\psi$ and
$\Upsilon$ polarizations have been specifically addressed in
Refs.~\cite{bib:ImprovedPolarization} and~\cite{big:Pol4Exps}.
Here we show that the frame-invariant relation remains valid for any $J=1$
particle decaying into a fermion-antifermion pair, even in the presence of
parity-violating effects.  Besides being useful in studies of the properties of
$W$ and $Z$ bosons, this relation can be particularly relevant for the
interpretation of new dilepton signals, where the production mechanisms and
polarizations of the candidate bosons are, a priori, unknown.
Besides improving the measurement of the spin alignment of the decaying
particle, the adoption of a frame-independent perspective also provides
more significant measurements of the parity asymmetry of the decay.
Furthermore, with respect to a
multidimensional analysis determining the frame-dependent angular coefficients
and their correlated experimental errors, the rotation-invariant definitions of
spin alignment and of parity asymmetry also provide an easier way of assessing the
significance of the global anisotropy of the decay distributions. Finally,
frame-independent results are less sensitive to specific choices of
polarization frames or other experimental constraints.

\section{Decay angular distribution of vector particles in fermion pairs}

We start by considering the most general angular distribution of
the di-fermion decay, expressed keeping track of the average angular
momentum composition of the decaying particle. First we address the case of a
single production ``subprocess'', denoted by the index $(i)$, in which the vector
particle $V$ is always formed as a specific superposition of the three $J_z$
eigenstates, with eigenvalues $m = +1, -1, 0$, with respect to a chosen
axis $z$,
\begin{equation}
  | V^{(i)} \rangle =  b^{(i)}_{+1} \,
  |\hspace{-.2em}+\hspace{-.2em}1\rangle + b^{(i)}_{-1} \,
  |\hspace{-.2em}-\hspace{-.2em}1\rangle +
  b^{(i)}_{0} \, |0\rangle \, . \label{eq:state}
\end{equation}
The decay angular distribution is described in the $V$ rest frame, where the
common direction of the two leptons define the reference axis $z^\prime$.
We denote by
$\vartheta$ and $\varphi$ the (polar and azimuthal) angles formed by one of the
two fermions with, respectively, the polarization axis $z$ and the $xz$ plane.
For a sufficiently massive decaying particle, the fermions can be considered
massless and helicity is conserved. The $| V^{(i)} \rangle$ decay
angular distribution is, then,
\begin{align}
  & W^{(i)}(\cos \vartheta, \varphi) \, \propto \,
  \frac{\mathcal{N}^{(i)}}{(3 + \lambda^{(i)}_{\vartheta})} \,
  (1  + \lambda^{(i)}_{\vartheta} \cos^2 \vartheta\  +
  \nonumber
\end{align}
\begin{align}
  & + \lambda^{(i)}_{\varphi} \sin^2 \vartheta \cos 2 \varphi
  + \lambda^{(i)}_{\vartheta \varphi} \sin 2 \vartheta \cos \varphi  \label{eq:ang_distr_subproc_parvio}
   \\
  & + \lambda^{\bot (i)}_{\varphi} \sin^2 \vartheta \sin 2 \varphi +
  \lambda^{\bot (i)}_{\vartheta \varphi} \sin 2 \vartheta \sin \varphi \nonumber \\
  & + 2A^{(i)}_{\vartheta} \cos \vartheta
    + 2A^{(i)}_{\varphi} \sin \vartheta \cos \varphi
    + 2A^{\bot (i)}_{\varphi} \sin \vartheta \sin \varphi
  ) \, . \nonumber
\end{align}
The subprocess coefficients depend on the partial amplitudes, $a_{m
\kappa}^{(i)}$, corresponding to specific configurations of the decaying state
and of the two-fermion system,
\begin{align}
\begin{split}
  \lambda^{(i)}_{\vartheta} =
  & \, 1/ \mathcal{D}^{(i)} \, [\mathcal{N}^{(i)}-3
( |a_{0,+1}^{(i)}|^2 + |a_{0,-1}^{(i)}|^2 )] \, , \\[0.5mm]
  \lambda^{(i)}_{\varphi} =
  & \, 2/\mathcal{D}^{(i)} \, \mathrm{Re} (a_{+1,+1}^{(i)*}
  a_{-1,+1}^{(i)} + a_{+1,-1}^{(i)*} a_{-1,-1}^{(i)}) \, , \\[0.5mm]
    \lambda^{(i)}_{\vartheta \varphi} =
  & \, \sqrt{2}/\mathcal{D}^{(i)} \, \mathrm{Re} [ a_{0,+1}^{(i)*} ( a_{+1,+1}^{(i)} -
  a_{-1,+1}^{(i)}) \\[0.5mm]
  & \qquad + a_{0,-1}^{(i)*} ( a_{+1,-1}^{(i)} - a_{-1,-1}^{(i)})]  \, , \\[0.5mm]
  \lambda^{\bot (i)}_{\varphi} =
  & \, 2/\mathcal{D}^{(i)} \, \mathrm{Im} (a_{+1,+1}^{(i)*}
  a_{-1,+1}^{(i)} + a_{+1,-1}^{(i)*} a_{-1,-1}^{(i)}) \, , \\[0.5mm]
  \lambda^{\bot (i)}_{\vartheta \varphi} =
  & - \sqrt{2}/\mathcal{D}^{(i)} \, \mathrm{Im} [ a_{0,+1}^{(i)*} (
  a_{+1,+1}^{(i)} + a_{-1,+1}^{(i)}) \\[0.5mm]
  & \qquad + a_{0,-1}^{(i)*} (  a_{+1,-1}^{(i)} + a_{-1,-1}^{(i)})] \, , \\[0.5mm]
  A^{(i)}_{\vartheta} =
  & \, 1/\mathcal{D}^{(i)} \, ( |a_{+1,+1}^{(i)}|^2 + |a_{-1,-1}^{(i)}|^2 \\[0.5mm]
  & \qquad -|a_{+1,-1}^{(i)}|^2 - |a_{-1,+1}^{(i)}|^2 ) \, , \\[0.5mm]
  A^{(i)}_{\varphi} =
  & \, \sqrt{2} /(2 \mathcal{D}^{(i)}) \,
  \mathrm{Re} [ a_{0,+1}^{(i)*} ( a_{+1,+1}^{(i)} + a_{-1,+1}^{(i)}) \\[0.5mm]
  & \qquad - a_{0,-1}^{(i)*} (a_{+1,-1}^{(i)} + a_{-1,-1}^{(i)})]  \, , \\[0.5mm]
  A^{\bot (i)}_{\varphi} =
  & - \sqrt{2} /(2 \mathcal{D}^{(i)}) \,
  \mathrm{Im} [ a_{0,+1}^{(i)*} ( a_{+1,+1}^{(i)} - a_{-1,+1}^{(i)}) \\[0.5mm]
  & \qquad - a_{0,-1}^{(i)*} ( a_{+1,-1}^{(i)} - a_{-1,-1}^{(i)})] \, ,
  \label{eq:lambdas_vs_amplitudes_parvio}
\end{split}
\end{align}
where $m$ is the $J_z$ component of $V$, $\kappa
= \pm 1$ is the total angular momentum projection of the two-fermion system along
the fermion momentum direction in the $V$ rest fame,
$\mathcal{N}^{(i)} = \, \sum_{m,\kappa} |a_{m\kappa}^{(i)}|^2$, and
$\mathcal{D}^{(i)} = \mathcal{N}^{(i)} + |a_{0,+1}^{(i)}|^2 +
|a_{0,-1}^{(i)}|^2$. In the special case of parity-conserving decays
$a_{m, +1}^{(i)} = a_{m, -1}^{(i)}$ and, hence,
$A^{(i)}_{\vartheta} = A^{(i)}_{\varphi} = A^{\bot(i)}_{\varphi} = 0$.

In the presence of $n$ contributing production processes with weights
$f^{(i)}$, the resulting observable distribution has a general expression
formally analogous to Eq.~\ref{eq:ang_distr_subproc_parvio}:
\begin{align}
 & W(\cos \vartheta, \varphi) \,
\propto \, \frac{1}{(3 + \lambda_{\vartheta})} \,
  (1 + \lambda_{\vartheta} \cos^2 \vartheta \nonumber \\[0.4mm]
& + \lambda_{\varphi} \sin^2 \vartheta \cos 2 \varphi +
  \lambda_{\vartheta \varphi} \sin 2 \vartheta \cos \varphi
\label{eq:ang_distr_general} \\
& + \lambda^{\bot}_{\varphi} \sin^2 \vartheta \sin 2 \varphi +
  \lambda^{\bot}_{\vartheta \varphi} \sin 2 \vartheta \sin \varphi
\nonumber \\
& + 2A_{\vartheta} \cos \vartheta
    + 2A_{\varphi} \sin \vartheta \cos \varphi
    + 2A^{\bot}_{\varphi} \sin \vartheta \sin \varphi
) \, . \nonumber
\end{align}
The observable coefficients, $X = \lambda_{\vartheta}$,
$\lambda_{\varphi}$, $\lambda_{\vartheta
  \varphi}$, $\lambda^{\bot}_{\varphi}$, $\lambda^{\bot}_{\vartheta \varphi}$,
  $A_{\vartheta}$, $A_{\varphi}$ and $A^{\bot}_{\varphi}$,
are weighted averages of the corresponding single-subprocess parameters,
$X^{(i)}$,
\begin{equation}
  X \, = \, \sum_{i = 1}^{n} \frac{f^{(i)} \mathcal{N}^{(i)}}{3 +
    \lambda_{\vartheta}^{(i)}} \,
  X^{(i)} \left/ \sum_{i = 1}^{n} \frac{f^{(i)} \mathcal{N}^{(i)}}{3 +
      \lambda_{\vartheta}^{(i)}} \right. \, .
\label{eq:parameters}
\end{equation}
While the most general form of the decay distribution is always given by
Eq.~\ref{eq:ang_distr_general} for any choice of polarization frame, the
coefficients depend on this choice. In this paper we only consider inclusive
production and, therefore, the only sensible experimental definition of the $xz$
plane is the production plane, containing the directions of the colliding
particles and of the decaying particle itself. A change of polarization frame
is thus a rotation in the production plane, parametrized by one angle,
$\delta$. The corresponding transformations of the coefficients are
\begin{equation}
\begin{split}
\lambda^{\prime}_\vartheta & =  \frac{\lambda_\vartheta
 - 3\Lambda }{1 + \Lambda }\, ,  \quad
\lambda^{\prime}_\varphi  =  \frac{ \lambda_\varphi
 + \Lambda }{1 + \Lambda }\, ,  \\
\lambda^{\prime}_{\vartheta \varphi} & =  \frac{ \lambda_{\vartheta \varphi}
\cos 2\delta  - \frac{1}{2}\, (\lambda_\vartheta - \lambda_\varphi )
\sin 2\delta } { 1 + \Lambda }\, , \\
\lambda^{\bot \prime}_\varphi  & = \frac{ \lambda^{\bot}_\varphi \cos \delta -
\lambda^{\bot}_{\vartheta \varphi} \sin \delta }
{ 1 + \Lambda }\, ,  \\
\lambda^{\bot \prime}_{\vartheta \varphi}  & = \frac{ \lambda^{\bot}_\varphi
\sin \delta + \lambda^{\bot}_{\vartheta \varphi} \cos \delta }
{ 1 + \Lambda }\, ,  \\
A^{\prime}_\vartheta  & = \frac{ A_\vartheta \cos \delta + A_{\varphi} \sin
\delta }{ 1 + \Lambda }\, ,  \\
A^{\prime}_\varphi  & = \frac{ -A_\vartheta \sin \delta + A_{\varphi} \cos
\delta }{ 1 + \Lambda }\, ,  \quad
A^{\bot \prime}_\varphi  = \frac{ A^{\bot}_\varphi }{ 1 + \Lambda }\, ,  \\
\mathrm{with} \quad \Lambda & = \frac{1}{2}\, (\lambda_\vartheta -
\lambda_\varphi)\sin^2 \delta - \frac{1}{2}\, \lambda_{\vartheta \varphi} \sin
2\delta  \, . \label{eq:coefficients_transf}
\end{split}
\end{equation}

In this paper we focus on quantities which are invariant under this
transformation. In fact, from Eq.~\ref{eq:coefficients_transf} it is immediate
to derive the following covariance relations:
\begin{equation}
\begin{split}
3 + \lambda^{\prime}_\vartheta & =  \frac{3 + \lambda_\vartheta }{1 + \Lambda
}\, ,  \\
1 - \lambda^{\prime}_\varphi & =  \frac{ 1 - \lambda_\varphi }{1 + \Lambda }\, ,  \\
\sqrt{\lambda^{\bot \prime 2}_\varphi + \lambda^{\bot \prime 2}_{\vartheta
\varphi} } & = \frac{ \sqrt{\lambda^{\bot 2}_\varphi + \lambda^{\bot
2}_{\vartheta \varphi} } }{1 + \Lambda }\, ,  \\
\sqrt{ A^{\prime 2}_\vartheta + A^{\prime 2}_\varphi + A^{\bot \prime
2}_{\varphi} } & = \frac{ \sqrt{ A^{2}_\vartheta + A^{2}_\varphi
+ A^{\bot 2}_{\varphi} } }{1 + \Lambda }\, ,  \\
\sqrt{ A^{\prime 2}_\vartheta + A^{\prime 2}_\varphi } & = \frac{ \sqrt{
A^{2}_\vartheta + A^{2}_\varphi } }{1 + \Lambda }\, . \label{eq:covariant
quantities}
\end{split}
\end{equation}
Naturally, relations linearly depending on these are also covariant.
The ratio of any two
of the covariant quantities defined by these relations is independent of the
chosen polarization frame. Certain specific
frame-independent quantities are discussed in the following sections.

In the absence of parity-violating processes, the coefficients
$A_{\vartheta}$, $A_{\varphi}$ and $A^{\bot}_{\varphi}$ are zero.
Besides, the observed \emph{event
distribution} must be symmetric by reflection with respect to the production
plane, even if the ``natural'' polarization plane does not coincide
\emph{event-by-event} with the production plane.  Therefore,
also $\lambda^{\bot}_{\varphi}$ and
$\lambda^{\bot}_{\vartheta \varphi}$, multiplying terms that are asymmetric by
the reflection transformation $\varphi \rightarrow \pi + \varphi$, are
unobservable, because they vanish on average.
In the more general case when parity is not conserved, as in $Z / \gamma^{*}\to
\ell\ell$ and $W\to\ell\nu$, event-by-event topological asymmetries with
respect to the experimental production plane can result in non-zero values of
the observed $\lambda^{\bot}_{\varphi}$ and $\lambda^{\bot}_{\vartheta
\varphi}$, even if these coefficients, as well as $A^{\bot}_{\varphi}$, are expected to
remain negligible because of the kinematic averaging.
In summary, in inclusive-production studies the observable di-fermion
distribution is essentially described by five coefficients: the three
``spin-alignment'' parameters $\lambda_{\vartheta}$, $\lambda_{\varphi}$ and
$\lambda_{\vartheta \varphi}$, containing information about the average angular
momentum composition of the decaying particle, and the two ``parity-asymmetry''
parameters $A_{\vartheta}$ and $A_{\varphi}$, expressing the parity properties
of the decay.

The parity-conserving part of the distribution can be written as a quadratic
form in the momentum coordinates $p_x, p_y, p_z$ of the lepton,
\begin{align}
\begin{split}
  1 + \lambda_{\vartheta} \cos^2 \vartheta
    + \lambda_{\varphi} \sin^2 \vartheta \cos 2 \varphi
    + \lambda_{\vartheta \varphi} \sin 2 \vartheta \cos \varphi \\
\; \propto \;
\begin{pmatrix} p_x & p_y & p_z \end{pmatrix}
\begin{pmatrix} 1 + \lambda_\varphi & 0 & \lambda_{\vartheta \varphi} \\
                       0    & 1 - \lambda_\varphi &     0       \\
                \lambda_{\vartheta \varphi} & 0 & 1 + \lambda_\vartheta \\
\end{pmatrix}
\begin{pmatrix} p_x \\ p_y \\ p_z \end{pmatrix}
  \, , \label{eq:ang_distr_quadratic_form}
\end{split}
\end{align}
showing that the $\lambda_{\vartheta \varphi}$ term is non-zero when
the axes of the reference frame do not coincide with the principal axes of
symmetry of the parity-conserving distribution. There always exists a
polarization frame where $\lambda_{\vartheta \varphi} = 0$ and the
parity-conserving part of the decay distribution is ``diagonal''. In this
frame, the shape of the decay distribution is univocally described by the parameters
$\lambda_{\vartheta}$, $\lambda_{\varphi}$, $A_{\vartheta}$ and $A_{\varphi}$.
These parameters are always bound, in any frame, between $-1$ and
$+1$, while $|\lambda_{\vartheta \varphi}| \le \sqrt{2}/2$.

\section{Rotation-invariant polarization observable}

The rotation-covariance properties of the generic $J=1$ state defined in
Eq.~\ref{eq:state} imply that each amplitude combination
$b_{+1}^{(i)}+b_{-1}^{(i)}$ is invariant by rotation about the $y$
axis~\cite{bib:LamTung}. In terms of decay amplitudes, this means that the
combinations $a_{+1,\kappa}^{(i)} + a_{-1,\kappa}^{(i)}$ (with $\kappa = \pm
1$) are invariant by rotation about $y$. The ``normalization'' $\sum_{m,\kappa}
|a_{m\kappa}^{(i)}|^2$ is obviously invariant under any rotation. In inclusive
production studies, all experimentally definable polarization axes belong to the
production plane, implying that a change of polarization frame means a rotation about
$y$. Hence, for each subprocess, the quantity
\begin{equation}
\mathcal{F}^{(i)}=\frac{|a_{+1,+1}^{(i)}+a_{-1,+1}^{(i)}|^2 +
|a_{+1,-1}^{(i)}+a_{-1,-1}^{(i)}|^2}{2\, \sum_{m,\kappa} |a_{m\kappa}^{(i)}|^2
} \,  \label{eq:kappa_i_parvio}
\end{equation}
(bound between 0 and 1) is independent of the chosen polarization frame.
Using also Eqs.~\ref{eq:lambdas_vs_amplitudes_parvio} and~\ref{eq:parameters},
we find that the combination of observable parameters
\begin{equation}
  \mathcal{F} \, = \, \frac{ \sum_{i = 1}^{n} f^{(i)}
    \mathcal{N}^{(i)} \mathcal{F}^{(i)}}{\sum_{i =
      1}^{n} f^{(i)} \mathcal{N}^{(i)}}  \, = \, \frac{1 +
    \lambda_\vartheta + 2
    \lambda_\varphi}{3 + \lambda_\vartheta} \label{eq:F}
\end{equation}
is frame-invariant.
This observable invariant polarization parameter, $\mathcal{F}$, is
the average of the corresponding sub-process invariants, calculated irrespectively of
the directions of the $n$ polarization axes.

For example, $\mathcal{F} = 0$ when all sub-process distributions are of the
kind $1 - \cos^2 \vartheta^{(i)}$, with respect to $n$ specific polarization
axes $z^{(i)}$ (belonging to the production plane). When the individual
distributions have the shapes $1 + \cos^2 \vartheta^{(i)}$, $\mathcal{F} =
1/2$. When the observed distribution is isotropic, $\mathcal{F} = 1/3$. We note
that $\mathcal{F}$ does not depend on the spin-alignment parameter
$\lambda_{\vartheta \varphi}$. In fact, while $\lambda_{\vartheta \varphi}$ can
be interpreted as a measure of the ``tilt'' of the chosen system of axes with
respect to the principal axes of symmetry of the distribution, $\mathcal{F}$
represents an intrinsic (rotation-independent) characteristic of the
\emph{shape} of the angular distribution.

The existence of a frame-independent polarization observable is a completely
general result, valid for the decay of any vector particle into fermion pairs,
irrespectively of its production mechanism.

\begin{figure}[h]
\centering
\includegraphics[width=\linewidth]{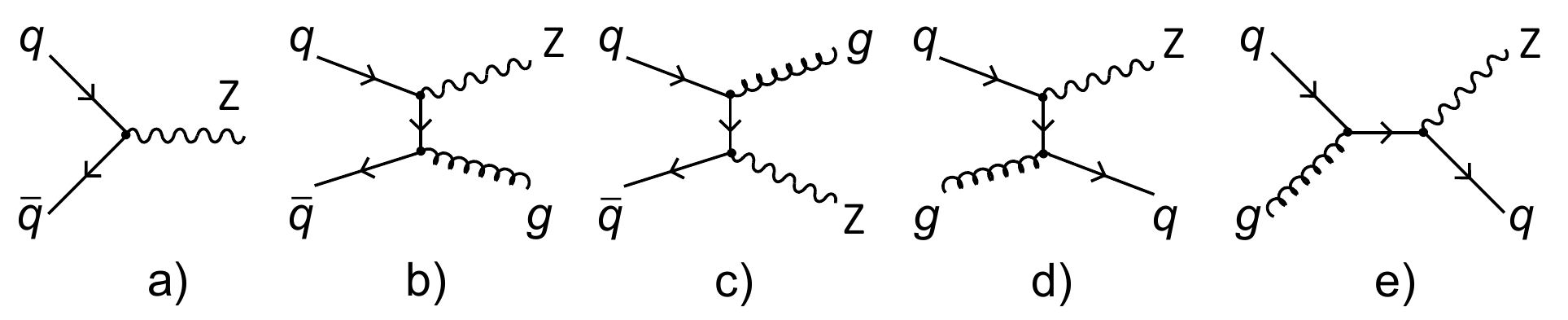}
\caption{$O(\alpha_s^0)$ and $O(\alpha_s^1)$ processes for $Z/\gamma^{*}$ and
$W$ production, giving rise to transverse dilepton polarizations along
different quantization axes: Collins--Soper (a), Gottfried--Jackson (b, c, d) and
helicity (e).} \label{fig:diagrams}
\end{figure}

As a special case, it is easy to recognize that Eq.~\ref{eq:F} resembles the
``Lam-Tung relation''~\cite{bib:LamTung}, $\lambda_\vartheta \, + \, 4 \,
\lambda_\varphi \, = \, 1$,  a result valid for Drell-Yan production in
perturbative QCD. This relation is trivially derivable when the invariants
$\mathcal{F}^{(i)}$ (and, thus, $\mathcal{F}$) are equal to $1/2$. The Lam-Tung
relation can therefore be reinterpreted as a consequence of rotational
invariance and of the fact that all contributing processes produce transversely
polarized lepton pairs. The latter ingredient can in turn be inferred from
helicity conservation and, thus, the Lam-Tung relation can be extended to the
description of direct $Z/\gamma^{*}$ and $W$ production.
Figure~\ref{fig:diagrams} shows the processes representing the $O(\alpha_s^0)$
and $O(\alpha_s^1)$ contributions to direct $Z/\gamma^{*}$ and $W$ production.
The assumption of helicity conservation at each diquark-boson vertex leads in
each case to the prediction of transverse polarizations, even if with respect
to different axes: Collins--Soper (CS)~\cite{bib:coll_sop}, 
approximation (neglecting the parton intrinsic transverse momentum)
of the direction of the colliding quark and antiquark (Fig.~\ref{fig:diagrams}\,a),
Gottfried--Jackson (GJ)~\cite{bib:gott_jack}, approximation of the direction of 
the single quark or antiquark coupling to the vector boson (Fig.~\ref{fig:diagrams}\,b,c,d), 
and helicity, direction of the vector boson momentum (Fig.~\ref{fig:diagrams}\,e).

It can be convenient to consider the following alternative
frame-independent observable:
\begin{equation}
  \tilde{\lambda} \, \equiv \, \frac{3\mathcal{F}-1}{1-\mathcal{F}}
  \, = \, \frac{\lambda_\vartheta + 3 \lambda_\varphi }{1 - \lambda_\varphi} \, .
    \label{eq:lambda_tilde}
\end{equation}
In the special case when the observed distribution is the superposition of $n$
``elementary'' distributions of the kind $1 + \lambda_\vartheta^{(i)}
\cos^2\vartheta$ with respect to $n$ different polarization axes,
$\tilde{\lambda}$ represents a weighted average of the $n$ polarizations, made
irrespectively of the orientations of the corresponding axes,
\begin{equation}
 \tilde{\lambda} \; = \;
\sum_{i = 1}^{n} \frac{f^{(i)}}{3 + \lambda_{\vartheta}^{(i)}} \,
\lambda^{(i)}_{\vartheta} \, \bigg/ \, \sum_{i = 1}^{n} \frac{f^{(i)}}{3 +
  \lambda_{\vartheta}^{(i)}} \, ,
\label{eq:lambda_tilde_meaning}
\end{equation}
with $\tilde{\lambda} = +1$ when all contributing processes have natural
transverse polarizations and $\tilde{\lambda} = -1$ when they are all
longitudinal.

\begin{figure}[htb]
\centering
\includegraphics[width=0.9\linewidth]{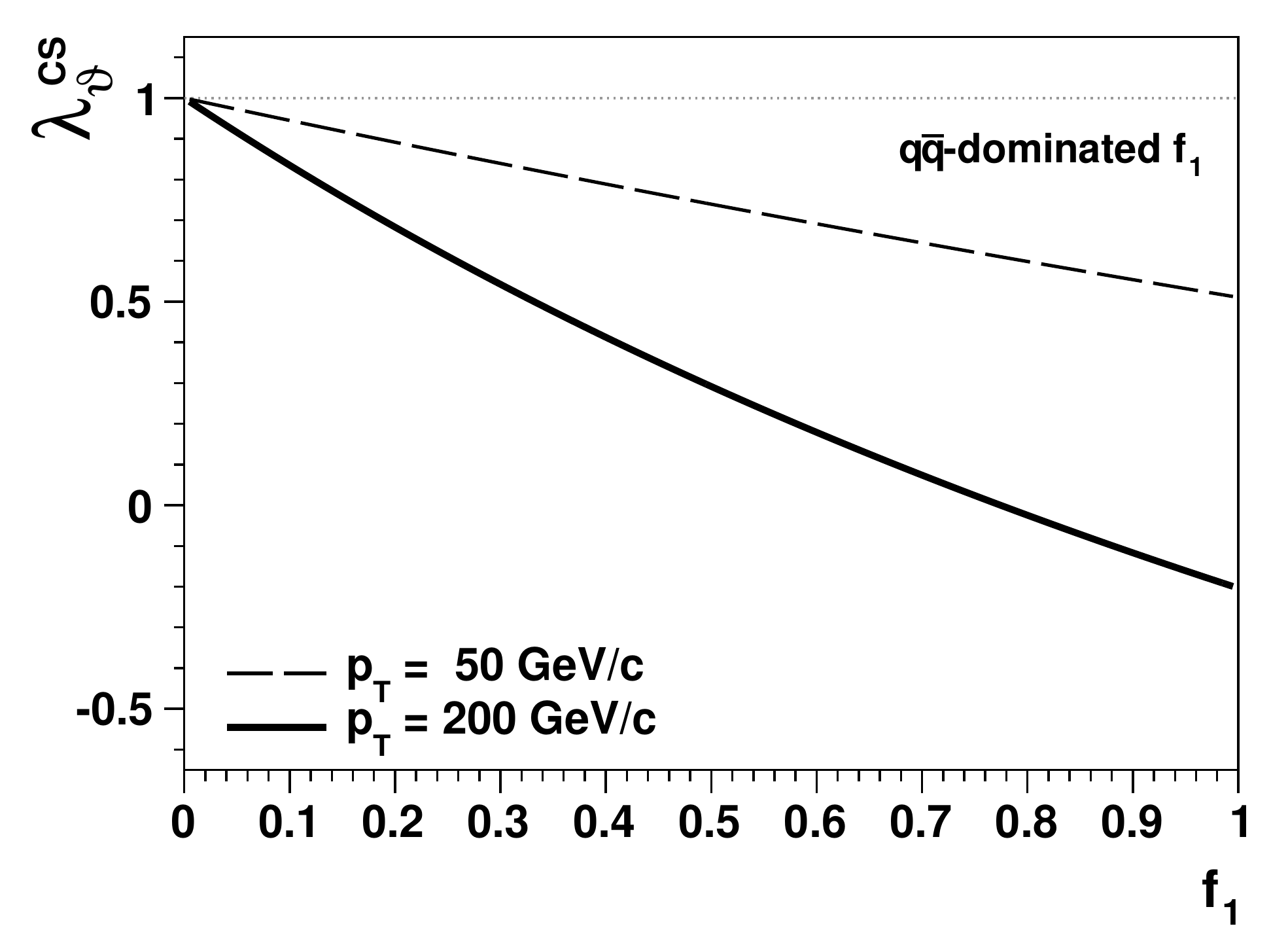}\\
\includegraphics[width=0.9\linewidth]{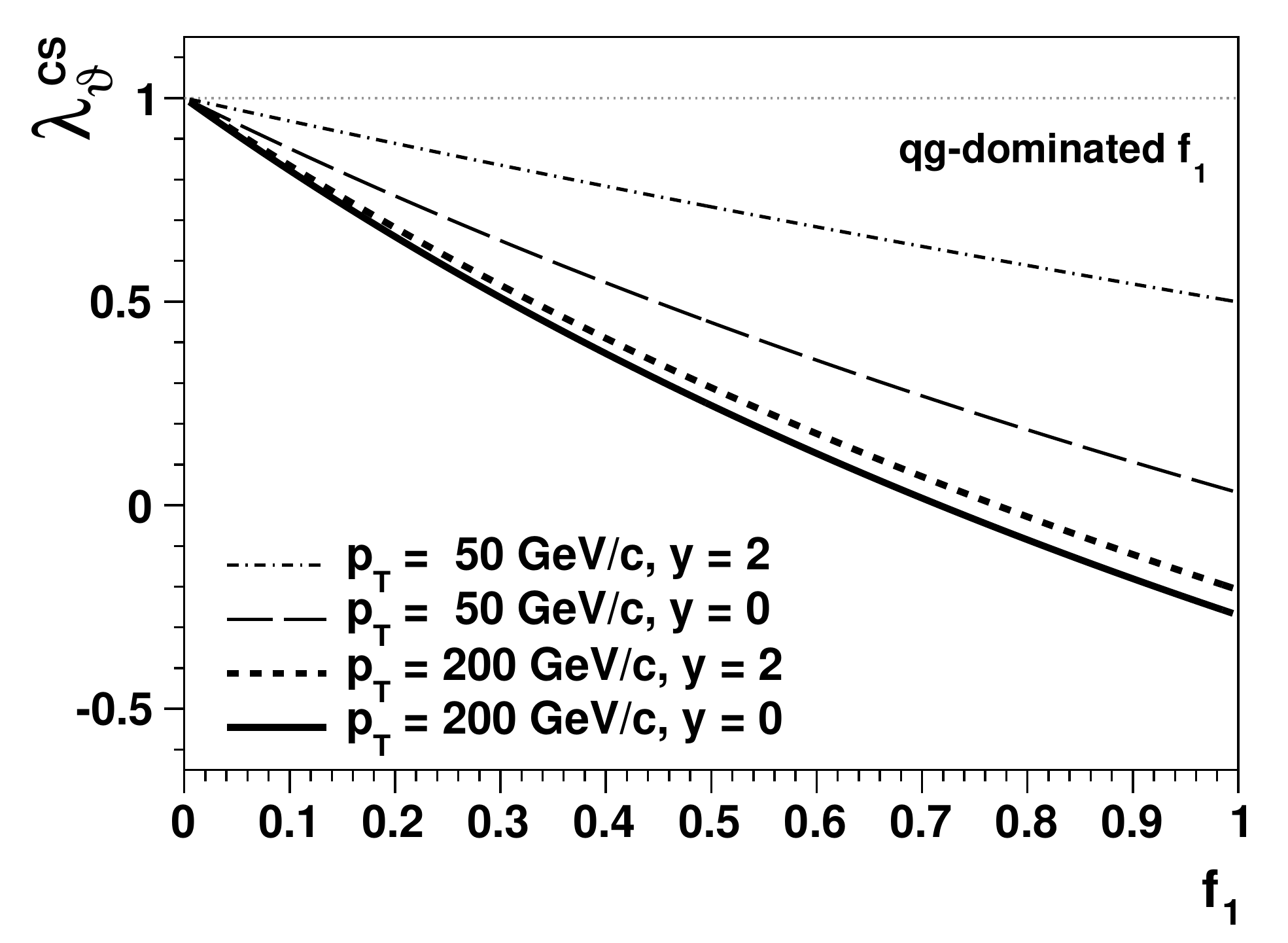}
\caption{Polar anisotropy of $W$ decay, observable in the Collins--Soper frame,
as a function of the relative contribution of $O(\alpha_s^1)$ processes, $f_1$, for
different $p_\mathrm{T}$ and rapidity values. The $O(\alpha_s^1)$ contributions
are assumed to be dominated by the quark-antiquark diagrams
(Fig.~\ref{fig:diagrams}\,b-c) in the top panel and by the quark-gluon diagrams
(Fig.~\ref{fig:diagrams}\,d-e) in the bottom panel. In the top panel 
only two curves are drawn because $\lambda^{\rm CS}_{\vartheta}$ 
is rapidity-independent in the case of quark-antiquark dominance.} 
\label{fig:W_vs_f1}
\end{figure}

In the case of direct $Z/\gamma^{*}$ or $W$ production, for example, the polar
anisotropy observed in the Collins--Soper frame is maximal only at
$p_\mathrm{T}=0$ (when all frames are degenerate). With increasing
$p_\mathrm{T}$, the polarization axes become increasingly different from each
other and the processes represented in Fig.~\ref{fig:diagrams} will add up to a
significantly reduced observable polarization.
Figure~\ref{fig:W_vs_f1} shows how this reduction of $\lambda^{\rm
CS}_{\vartheta}$, for high $p_\mathrm{T}$ $W$ production, is affected by the
relative contribution of the sub-processes transversely polarized in the GJ and
helicity frames.
The measurement of $\tilde{\lambda}$ always yields, instead, the value $+1$,
directly reflecting the intrinsic, fully transverse polarization characteristic
of the produced $W$.

On the other hand, the measurement of the frame-dependent parameters can
provide further insight into the production processes.
For example, from the
comparison of the top and bottom panels of Fig.~\ref{fig:W_vs_f1} we deduce
that accurate rapidity-dependent measurements of $\lambda^{\rm
CS}_{\vartheta}$, made at not too high $p_\mathrm{T}$, can reveal if the QCD
corrections are dominated by quark-antiquark or by quark-gluon diagrams, 
given that only in the latter case $\lambda^{\rm CS}_{\vartheta}$ would depend 
on rapidity.

\section{Rotation-invariant parity-violating asymmetry}
\label{sec:parvio}

We will now study the rotational properties of the parity-violating coefficients
$A_{\vartheta}$, $A_{\varphi}$ and $A^{\bot}_{\varphi}$.  We start by noticing that,
for each subprocess, the combination
\begin{equation}
  \tilde{\mathcal{A}}^{(i)}  \; = \; \frac{4}{3 +
    \lambda^{(i)}_\vartheta} \, \sqrt{A^{(i)2}_\vartheta
  + A^{(i)2}_\varphi + A^{\bot(i)2}_\varphi } \, ,
  \label{eq:frame_independent_asymmetry}
\end{equation}
taking values between 0 and 1, is invariant under \emph{any} rotation.
Furthermore, the rotational properties of the angular momentum eigenstate
$|V^{(i)} \rangle$ imply that there exists a quantization axis $z^{(i)\star}$
with respect to which $a_{0,+1}^{(i)\star} =
a_{0,-1}^{(i)\star}=0$~\cite{bib:LamTung}. Along this axis (from
Eq.~\ref{eq:lambdas_vs_amplitudes_parvio})
\begin{equation}
\lambda^{(i)\star}_\vartheta = + 1 \, , \quad A^{(i)\star}_\varphi = 0 \, , \quad
A^{\bot (i)\star}_{\varphi} = 0  \label{eq:maxAsymmAxis}
\end{equation}
and, moreover, $\lambda^{(i)\star}_{\vartheta \varphi} = \lambda^{\bot
  (i)\star}_{\vartheta \varphi} = 0$, while $A^{(i)\star}_\vartheta$,
$\lambda^{(i)\star}_\varphi$ and $\lambda^{\bot (i)\star}_{\varphi}$ can be
different from zero (in fact, $\lambda^{(i)\star}_\varphi =
2\,\mathcal{F}^{(i)}-1$, from Eq.~\ref{eq:F}). This means that each single
subprocess is characterized by its own ``natural'' axis, along which the
polarization is fully transverse and the entire magnitude of the
parity-violating effect is expressed by the coefficient
$A^{(i)\star}_\vartheta$,
\begin{equation}
  A^{(i)\star}_\vartheta = \;
  \frac{ |a_{+1,+1}^{(i)\star}|^2 + |a_{-1,-1}^{(i)\star}|^2
  -|a_{+1,-1}^{(i)\star}|^2 - |a_{-1,+1}^{(i)\star}|^2 }{
  |a_{+1,+1}^{(i)\star}|^2 + |a_{-1,-1}^{(i)\star}|^2
  +|a_{+1,-1}^{(i)\star}|^2 + |a_{-1,+1}^{(i)\star}|^2 }  \, .
  \label{eq:A_theta_natural_axis}
\end{equation}
Therefore, the axis-independent parity asymmetry
$\tilde{\mathcal{A}}^{(i)}$ represents the magnitude of
the \emph{maximum} parity-violating asymmetry, as would be measured if
it were possible to isolate the $i$-th single subprocess and choose
$z^{(i)\star}$ as polarization axis, $\tilde{\mathcal{A}}^{(i)} =
|A^{(i)\star}_\vartheta|$.

While the polarization information contained in the observable
parameter $\mathcal{F}$ is not smeared in the superposition of
subprocesses with different natural axes, smearing effects are
unavoidable in the observable parity asymmetry, as a consequence of
the intrinsic directionality induced by parity asymmetry.  To
quantify these effects, we define the asymmetry ``vectors''
\begin{equation}
  \vec{\mathcal{A}}^{(i)}  \; = \; \frac{4}{3 +
    \lambda^{(i)}_\vartheta} \, ( A^{(i)}_\vartheta ,
  A^{(i)}_\varphi , A^{\bot(i)}_\varphi ) \, ,
  \label{eq:vector_frame_independent_asymmetry}
\end{equation}
each one directed along the corresponding natural axis $z^{(i)\star}$ and equal
to $\tilde{\mathcal{A}}^{(i)}$ in magnitude. The observable rotation-invariant
quantity
\begin{equation}
  \tilde{\mathcal{A}}  \; = \;
  \frac{4}{3 + \lambda_\vartheta} \, \sqrt{A^{2}_\vartheta +
    A^{2}_\varphi + A^{\bot 2}_\varphi } \, ,
  \label{eq:observable_frame_independent_asymmetry_def}
\end{equation}
bound between 0 and~1, is equal to the following combination of the
elementary single-process asymmetries:
\begin{equation}
  \tilde{\mathcal{A}}  \; = \;
  \frac{ \sqrt{ \sum_{i,j = 1}^{n} f^{(i)} \mathcal{N}^{(i)} \vec{\mathcal{A}}^{(i)}
  \cdot f^{(j)} \mathcal{N}^{(j)} \vec{\mathcal{A}}^{(j)}  }  }
  {\sum_{i = 1}^{n} f^{(i)} \mathcal{N}^{(i)}}.
  \label{eq:observable_frame_independent_asymmetry_comb}
\end{equation}
The inequality
\begin{equation}
  \tilde{\mathcal{A}}  \; \le \;
  \frac{  \sum_{i = 1}^{n} f^{(i)} \mathcal{N}^{(i)} \tilde{\mathcal{A}}^{(i)} }
  {\sum_{i = 1}^{n} f^{(i)} \mathcal{N}^{(i)}}
  \label{eq:observable_frame_independent_asymmetry_upper_limit}
\end{equation}
expresses the fundamental observation that superimposed processes characterized
by different polarization axes partially cancel each other, leading to a less
significant measurable asymmetry, whatever the observation frame.  With this in
mind, Eq.~\ref{eq:observable_frame_independent_asymmetry_def} provides the
magnitude of the \emph{maximum observable} parity asymmetry, i.e.\ of the net
asymmetry as it can be measured along the polarization axis that maximizes it.

It may be experimentally convenient, when the detector itself does not
induce ``parity-violating'' acceptance effects on the
two decay fermions, to measure the following asymmetries:
\begin{align}
\begin{split}
\mathcal{A}_{\cos \vartheta} & \; = \; \frac{N(\cos \vartheta > 0) - N(\cos
\vartheta < 0)}{N(\cos \vartheta > 0) + N(\cos \vartheta < 0)} \;  = \;
\frac{3 A_\vartheta}{3+\lambda_\vartheta} \, , \\[2mm]
\mathcal{A}_{\cos \varphi} & \; = \;  \frac{N(\cos \varphi > 0) - N(\cos
\varphi < 0)}{N(\cos \varphi > 0) + N(\cos \varphi < 0)} \;   = \; \frac{3
A_\varphi}{3+\lambda_\vartheta} \, , \\[2mm]
\mathcal{A}_{\sin \varphi} & \; = \;  \frac{N(\sin \varphi > 0) - N(\sin
\varphi < 0)}{N(\sin \varphi > 0) + N(\sin \varphi < 0)} \;   = \; \frac{3
A^\bot_\varphi}{3+\lambda_\vartheta} \, .
\end{split} \label{eq:fb_asymmetries}
\end{align}
The first one is the so-called forward-backward asymmetry, $\mathcal{A}_{FB}$,
studied, for example, in experimental analyses of direct $Z/\gamma^{*}$ and $W$
production, and usually defined in the CS frame.
However, at high \pt, where processes beyond leading order contribute
significantly, the CS frame no longer closely reflects the topology of the
decay process. As discussed in the previous section, the polarization parameter
undergoes a strong reduction from its leading-order expectation for direct
production $\lambda_\vartheta^{\mathrm{CS}} = +1$. In a similar way, the polar
``projection'' of the asymmetry becomes smaller than the maximum observable
asymmetry. For example, among the processes of Fig.~\ref{fig:diagrams} for
direct $Z/\gamma^{*}$ and $W$ production, the quark-gluon diagrams (d) and (e)
are naturally polarized along the GJ and helicity axes, respectively, and
can lead, therefore, to observable azimuthal components of the parity asymmetry
in the CS frame.
Instead, the diagrams (b) and (c) do not affect $A_\varphi$ because
the vector boson is emitted either by the quark or by the antiquark and the two
cases cancel exactly their contribution.
%
%
The term $A_\varphi$ is especially important when the observed process has
significant contributions from sub-processes having a natural polarization
along the helicity axis.

For indirectly produced vector bosons (for example,
$W$ from top quark decay) the CS axis loses completely its role of optimal spin
quantization axis.
In such situations the significance of the measured parity-violating effect can
be improved, independently of the choice of the polarization frame, by
using a rotation-invariant combination of the \emph{three} asymmetries,
which coincides with the already defined $\tilde{\mathcal{A}}$:
\begin{equation}
  \tilde{\mathcal{A}} \; = \; \frac{4}{3} \, \sqrt{ \mathcal{A}^2_{\cos \vartheta}
  + \mathcal{A}^2_{\cos \varphi} + \mathcal{A}^2_{\sin \varphi} } \, .
  \label{eq:observable_frame_independent_asymmetry_vs_fb_asymmetries}
\end{equation}

The parameter $A^\bot_\varphi$ and the corresponding asymmetry
$\mathcal{A}_{\sin \varphi}$ should be small, as a consequence of the 
approximate symmetry with respect to the production plane expected for 
the decay distribution of inclusively produced vector bosons.  In any case,
these parameters can be neglected
without affecting the exactness of the frame-independent formalism.
In fact, the ``reduced'' invariant asymmetry
\begin{equation}
  \tilde{\mathcal{A}}_\mathrm{R} \; = \;  \frac{4}{3 +
    \lambda_\vartheta} \, \sqrt{A^{2}_\vartheta + A^{2}_\varphi }
  \; = \; \frac{4}{3} \, \sqrt{ \mathcal{A}^2_{\cos \vartheta}
    + \mathcal{A}^2_{\cos \varphi} }
  \label{eq:observable_frame_independent_asymmetry_reduced}
\end{equation}
is invariant under rotation about the $y$ axis (while
$\tilde{\mathcal{A}}$ is invariant under \emph{any} rotation) and is,
therefore, exactly independent of the choice of a polarization axis
belonging to the production plane.

\section{Experimental advantages of the rotation-invariant formalism}
\label{sec:advantages}

The rotation-invariant parameters $\mathcal{F}$ and $\tilde{\mathcal{A}}$ can
be determined directly from a fit of the angular distribution
(taking into account acceptance and efficiencies) with \emph{any} choice of
polarization frame, through a suitable substitution of parameters in
Eq.~\ref{eq:ang_distr_general}. For example:
\begin{equation}
\begin{split}
\lambda_\varphi & \, \rightarrow \, 1 - 1/2 (1-\mathcal{F})(3 +
\lambda_\vartheta) \, , \\
A_\vartheta & \, \rightarrow \, \frac{(3+\lambda_\vartheta)
\tilde{\mathcal{A}}}{4} \cos \xi \, , \\
A_\varphi & \, \rightarrow \, \frac{(3+\lambda_\vartheta)
\tilde{\mathcal{A}}}{4} \sin \xi \cos \zeta \, , \\
A^{\bot}_\varphi & \, \rightarrow \, \frac{(3+\lambda_\vartheta)
\tilde{\mathcal{A}}}{4} \sin \xi \sin \zeta \, ,
\label{eq:parameter_substitution}
\end{split}
\end{equation}
where, $A^{\bot}_\varphi$ being negligible, $\zeta$ may be set to
zero.

The determination of $\mathcal{F}$
and $\tilde{\mathcal{A}}$ has the advantage that these parameters
directly estimate the significance of the global anisotropy of the
distribution, while to derive the same information from the
frame-dependent coefficients we need to study the correlation between
$\lambda_\vartheta$ and $\lambda_\varphi$, and also between $A_\vartheta$ and
$A_\varphi$. This is especially important if the polarization, or parity asymmetry, is
not large, or if the frame-dependent parameters
are poorly determined because of insufficient statistics.

The kinematic dependence of the frame-dependent coefficients has, in general, a
``spurious'' component reflecting the choice of observation frame rather than
the intrinsic characteristics of the production processes, or how their mixture
changes with kinematics. This can be understood from
Eq.~\ref{eq:coefficients_transf}, $\delta$ explicitly depending on the momentum
of the produced particle~\cite{big:Pol4Exps}. For instance, a subprocess with a
``natural'' parity-asymmetry axis coinciding with the helicity axis would give
the contributions
\begin{align}
\begin{split}
  A_\vartheta^{(i)}  & =  \frac{ m \, p_\mathrm{L} \, p \, m_\mathrm{T} }{ p^2 \, m_\mathrm{T}^2 + \frac{1}{2} \,
    p_\mathrm{T}^2 \, E^2 \, (1-\lambda^{(i)\star}_\varphi) }  \,
    A^{(i)\star}_\vartheta \, , \\
  A_\varphi^{(i)}  & =  - \frac{ E \, p_\mathrm{T} \, p \, m_\mathrm{T} }{ p^2 \, m_\mathrm{T}^2 + \frac{1}{2} \,
    p_\mathrm{T}^2 \, E^2 \, (1-\lambda^{(i)\star}_\varphi) }  \,
    A^{(i)\star}_\vartheta
\label{eq:Atheta_transf_example}
\end{split}
\end{align}
to the parameters $A_\vartheta$ and $A_\varphi$ observed in the CS frame, with
$m$, $m_\mathrm{T}$, $E$, $p$, $p_\mathrm{T}$ and $p_\mathrm{L}$ being,
respectively, the mass, the transverse mass, the energy, and the total,
transverse and longitudinal momenta of the meson in the center-of-mass of the
collision. At sufficiently high $p_\mathrm{T}$,
$A^{(i)}_\vartheta$ vanishes and
$A^{(i)}_\varphi$ acquires the same significance of $A^{(i)\star}_\vartheta$.
Moreover, the strong, explicit dependence of $A^{(i)}_\vartheta$ and
$A^{(i)}_\varphi$ on the momentum of the particle implies a significant
dependence of the observed asymmetries on the experimental acceptance.
In general, such
misleading effects cannot be reduced with a suitable frame choice, being
impossible to eliminate in the presence of a mixture of processes
characterized by different natural axes. By definition, the frame-independent
polarization and asymmetry parameters are less sensitive (or even immune) to these
effects.
Their use improves the representation of the results and facilitates the
comparison between different experiments, and with theoretical
calculations~\cite{bib:ImprovedPolarization}.

\begin{figure}[ht]
\centering
\includegraphics[width=0.9\linewidth]{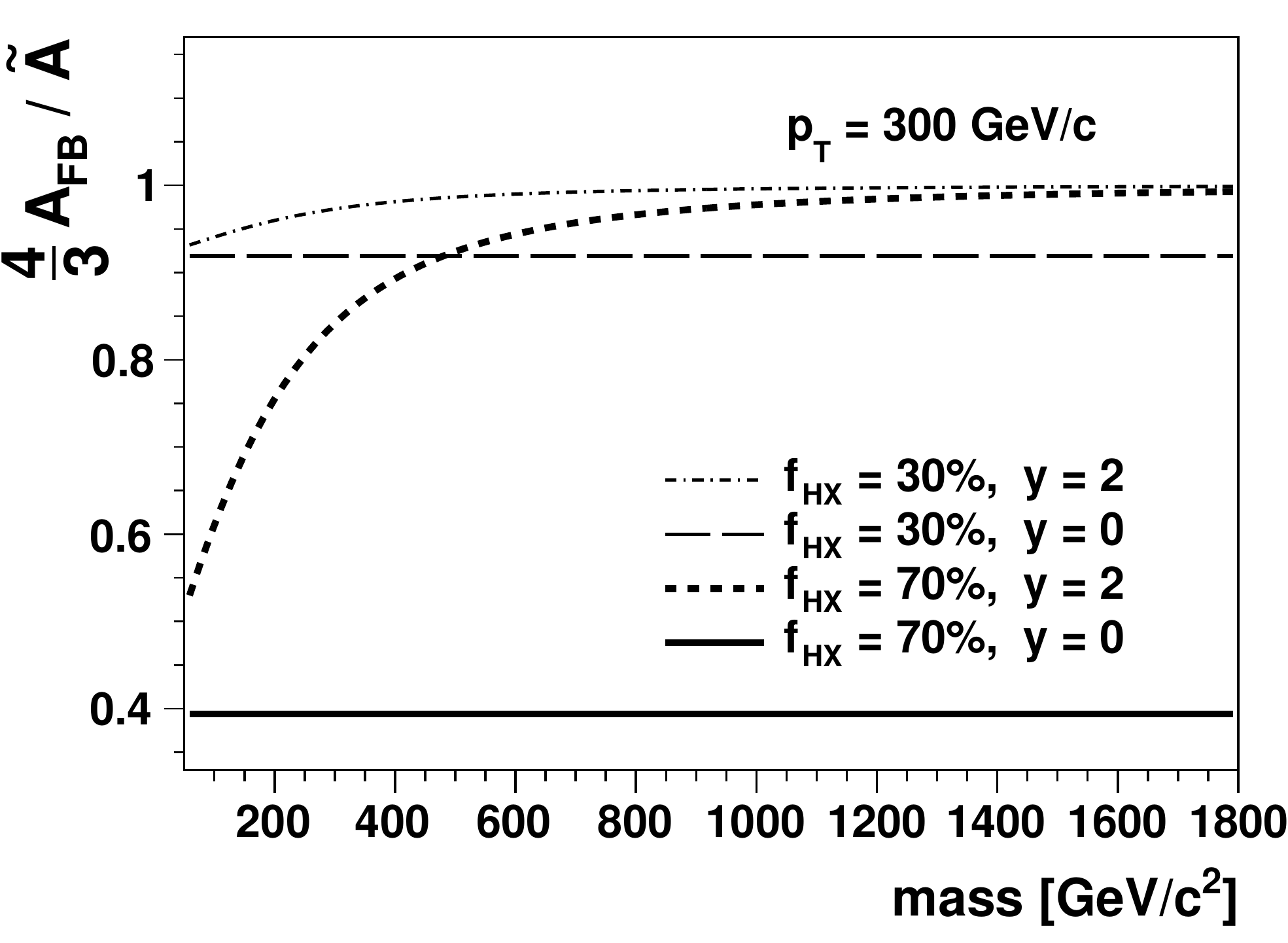}\\
\includegraphics[width=0.9\linewidth]{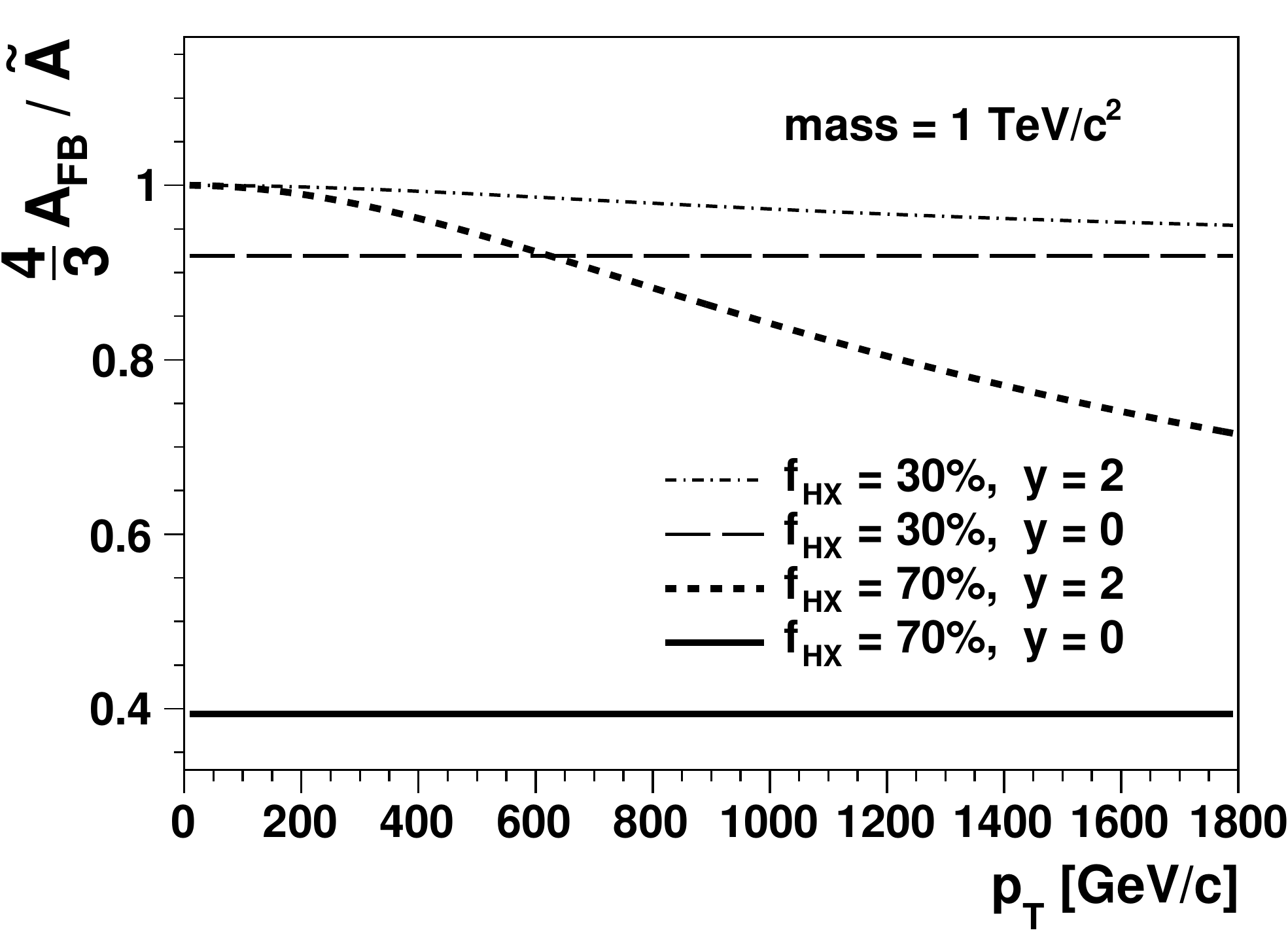}\\
\includegraphics[width=0.9\linewidth]{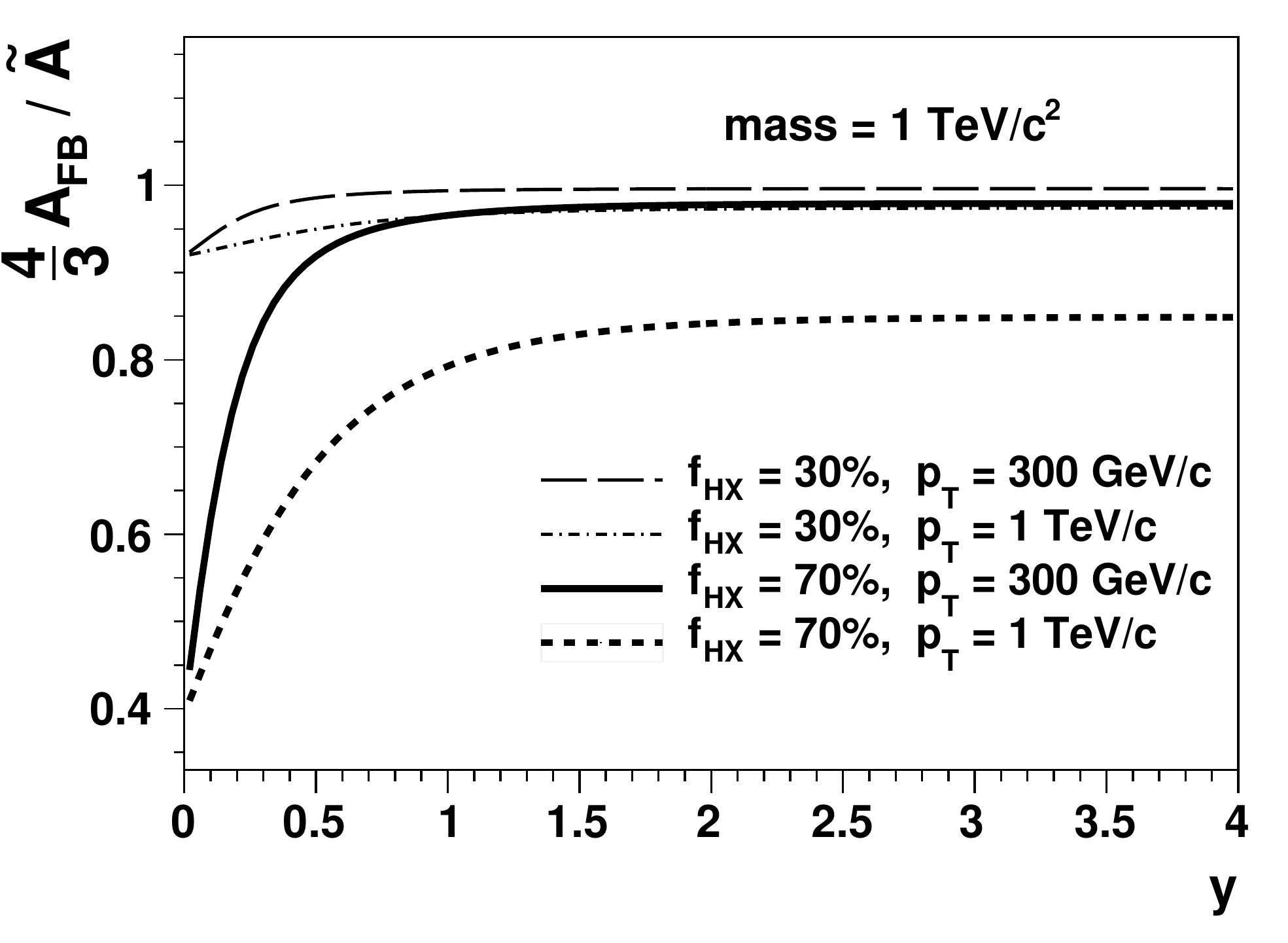}
\caption{Ratio between the forward-backward asymmetry (scaled up for
  consistency) and the maximally observable parity asymmetry, as a
  function of mass, $p_\mathrm{T}$ and rapidity of the vector boson, when
  a fraction $f_{\mathrm{HX}} = 30\%$ or $70\%$ of the cross section comes
  from processes naturally polarized in the helicity frame, the others
  being polarized in the CS frame.}
  \label{fig:kin}
\end{figure}

We will now motivate the use of a frame-independent evaluation of the
parity asymmetry by considering a concrete (albeit arbitrary)
scenario, where a heavy vector boson would be produced as a
superposition of two processes, one transversely polarized in the CS
frame and the other transversely polarized in the helicity frame.
This is illustrated in Fig.~\ref{fig:kin}. For simplicity of illustration, we
assume that the relative weight of these processes is independent of
kinematics. As long as the process naturally polarized in the CS frame
dominates, the use of the standard forward-backward asymmetry,
$\mathcal{A}_{FB}$, provides a relatively good approximation of the maximally
observable asymmetry, $\tilde{\mathcal{A}}$, apart from the trivial scale
factor 4/3. If, instead, a large fraction of the events are transversely
polarized in the helicity frame, the two approaches give very different
results.

Finally, the explicit verification of the frame-independence of the invariant
quantities represents a non-trivial check of the correctness of the
experimental analysis. In fact, acceptance limitations due to the
geometry of the detector and/or to its limited sensitivity to the
particle momenta act as strong kinematic cuts in the reconstructed
angular spectra and induce a fake polarization, which must be
accurately understood for a correct interpretation of the
measurements.  Such polarizing effects do not follow, in general, the
physical transformation rules of the decay distribution in the
particle rest frame, i.e., they do not obey the rotational-covariance
properties of the angular parameters.  Comparing frame-independent
results obtained in two significantly distinct frames can, thus,
reveal unaccounted systematic effects and lead to an 
improved evaluation of systematic uncertainties.

\section{Summary}

The spin alignment of a vector state decaying into fermion-antifermion (or
lepton-neutrino) pairs and inclusively observed in a given kinematic condition
is always described by a specific form of Eq.~\ref{eq:F}. This relation is
invariant with respect to rotations about the axis perpendicular to the 
production plane and provides, therefore, a frame-independent
definition of spin-alignment. Its existence derives from
basic rotational covariance properties of $J=1$ angular momentum eigenstates
and is a general result valid for parity-conserving and parity-violating
di-fermion decays.

The parity-violating terms of the decay distribution can be combined in a
second frame-independent relation, representing the magnitude of the maximum
observable parity asymmetry, as could be measured with the best possible choice
of quantization axis.

The adoption of the frame-invariant formalism for the measurement of
polarization and parity asymmetry presents several advantages in studies of the
decays of high-\pt\ bosons.
The rotation-invariant spin-alignment and parity-asymmetry parameters always
provide a representation of the maximum observable polarization and
parity-asymmetry effects. They are independent of assumptions on the production mechanisms
or on the optimal spin-quantization axis, and are immune to the effect of a possible ``unlucky
choice'' of the polarization frame. For example, with respect to the
usually adopted definition of the forward-backward asymmetry for $Z/\gamma^{*}$
decays, referred to one chosen polarization axis (typically the Collins--Soper
axis), the rotation-invariant formulation always gives a more significant
observable parity asymmetry.

By eliminating spurious kinematic dependencies induced by the
momentum-dependence of the angle between different physical quantization axes,
the use of frame-independent quantities also provides a more robust comparison
between different experiments, as well as between measurements and
calculations.
Additionally, the results can be cross-checked in two orthogonal frames,
probing the existence of unaccounted biases induced by experimental
limitations in the kinematic acceptance of the decay fermions.

\medskip

P.F., J.S.\ and H.K.W.\ acknowledge support from Funda\c{c}\~ao para a
Ci\^encia e a Tecnologia, Portugal, under contracts
SFRH/BPD/42343/2007, CERN/FP/ 109343/2009 and SFRH/BPD/42138/2007.


\end{document}